\documentclass[journal,comsoc]{IEEEtran}
\usepackage[T1]{fontenc}% optional T1 font encoding
\usepackage{color}
\usepackage{graphicx}
\usepackage[caption=false,font=footnotesize]{subfig}
\usepackage{fixltx2e}
\usepackage{tablefootnote}
\usepackage[table,x11names]{xcolor}
\usepackage{arydshln}

\ifCLASSINFOpdf
\else
\fi
\definecolor{dukeblue}{rgb}{0.0, 0.1, 0.9}
\definecolor{lightgreen}{rgb}{0.85, 0.9 , 0.89}
\definecolor{lightyellow}{rgb}{0.96, 0.91, 0.76}

\usepackage{amsmath}
\usepackage[cmintegrals]{newtxmath}
\begin{document}
\title{Deep Learning Framework for the Design of Orbital Angular Momentum Generators Enabled by Leaky-wave Holograms}

\author{Naser~Omrani,~Fardin~Ghorbani,~\IEEEmembership{Graduate~Student~Member,~IEEE,} ~Sina~Beyraghi, ~\IEEEmembership{Graduate~Student~Member,~IEEE},~Homayoon~Oraizi,~\IEEEmembership{Life~Senior,~IEEE},
        and~Hossein Soleimani% <-this % stops a space
\thanks{N. Omrani, F. Ghorbani, H. Oraizi and H. Soleimani are with the School
of Electrical Engineering, Iran University of Science and Technology, Tehran 1684613114,
Iran, e-mail: (h\_oraizi@iust.ac.ir).

S. Beyraghi is with the Department of Information and Communications Technologies, Pompeu Fabra University, Barcelona, Spain. He is also with Telefonica Company, Barcelona, Spain. He was with the School of Electrical Engineering, Iran University of Science and Technology, Tehran, Iran.}}


\markboth{}%
{Shell \MakeLowercase{\textit{et al.}}: Bare Demo of IEEEtran.cls for IEEE Communications Society Journals}



\maketitle

\begin{abstract}
	In this paper, we present a novel approach for the design of leaky-wave holographic antennas that generates OAM-carrying electromagnetic waves by combining Flat Optics (FO) and machine learning (ML) techniques. To improve the performance of our system, we use a machine learning technique to discover a mathematical function that can effectively control the entire radiation pattern, i.e., decrease the side lobe level (SLL) while simultaneously increasing the central null depth of the radiation pattern. Precise tuning of the parameters of the impedance equation based on holographic theory is necessary to achieve optimal results in a variety of scenarios. In this research, we applied machine learning to determine the approximate values of the parameters. We can determine the optimal values for each parameter, resulting in the desired radiation pattern, using a total of 77,000 generated datasets. 
    Furthermore, the use of ML not only saves time, but also yields more precise and accurate results than manual parameter tuning and conventional optimization methods.
\end{abstract}


\IEEEpeerreviewmaketitle

\section{Introduction}
\IEEEPARstart{I}{n} 
 addition to spin angular momentum (SAM) which represents the polarization state of electromagnetic (EM) waves, the orbital angular momentum (OAM) eigenstates as other intrinsic characteristics, which can be used to enhance the channel capacity of wireless telecommunication systems \cite{yan2014}.
The OAM waves contain an infinite number of orthogonal eigenstates, indicating the number of electromagnetic phase twists per wavelength around the phase singularity
\cite{allen1992}.
The phase singularity can be expressed mathematically by 
$exp(-jl\phi)$ term, in which $\phi$ represents the azimuth angle and $l$ is the topological charge of OAM waves.
Theoretically, OAM waves have null field magnitudes at their phase singularities, showing doughnut-like radiation patterns in which the divergence angle varies with the topological charge \cite{vallone2016}.
The above-mentioned features have attracted worldwide interest and form a cornerstone for prompting the applications of OAM waves \cite{chen2020}.
The OAM-carrying EM waves, so-called vortex beams, are propagation invariant localized waves that were first introduced by Allen \textit{et al.} \cite{allen1992} in 1992 for optical applications.
Since then, the multi-mode vortex beams, which provide a set of infinite orthogonal eigenstates with simultaneous transmit multiple independent data 
have been utilized for multiplexing/demultiplexing different signals \cite{tamburini2012}-\cite{ren2017}.
Hence, they have great potential to improve the channel capacity and spectral efficiency of communication systems in a wide range from radio frequencies (RF) to optics. 
Nowadays, machines as efficient tools are touching every aspect of human society. In the meantime, a mathematics-based interdisciplinary science combined with computer and biology science, called machine learning (ML), opened a new gate of very applicable science for researchers. ML accompanied by deep learning (DL) due to their powerful ability in data analysis, classification, and prediction has significant advantages in solving complex problems. Deep Neural Network (DNN) \cite{wan2023} exploits multi-layered artificial neural networks implemented with a computer to perform advanced tasks. 
Recently deep learning has been used in various fields, such as radar \cite{ ghorbani2023}, energy saving\cite{gomariz2023}, metasurface design \cite{ghorbani2021, ghorbani2021_2}, meta-analysis\cite{kempen2021}, signal processing \cite {ghorbani2020}, speech recognition \cite{saritha2023}, and optofluidic imaging \cite{siu2023}. In addition, deep learning has great potential to be utilized in communication \cite{zhang2023}. 
On the other hand, antenna performance enhancement has been a challenging task in communication from the past times until the present. To cope with this issue, researchers gain assistance from DL for antenna design \cite{khan2022,siddiqui2023},  which has been remarkably successful.

To the best of our knowledge, the application of a deep learning framework in the design of vortex beam leaky-wave holograms has not been reported in the literature.  However, in comparison to the other conventional methods, DL is characterized by high speed, high precision, and other characteristics that will be discussed in the following sections. In this scenario, we proposed and investigated a new framework based on Flat Optics (FO) \cite{minatti2016} and ML tools combined with mathematical theory to design holographic leaky-wave antennas (HLWA), generating OAM-carrying electromagnetic waves. To be more specific, we discovered a mathematical function for control over not only the side lobe level and central null depth, but also entire the radiation pattern. 

\begin{figure}
      \centering
      \includegraphics[width = 0.4\textwidth]{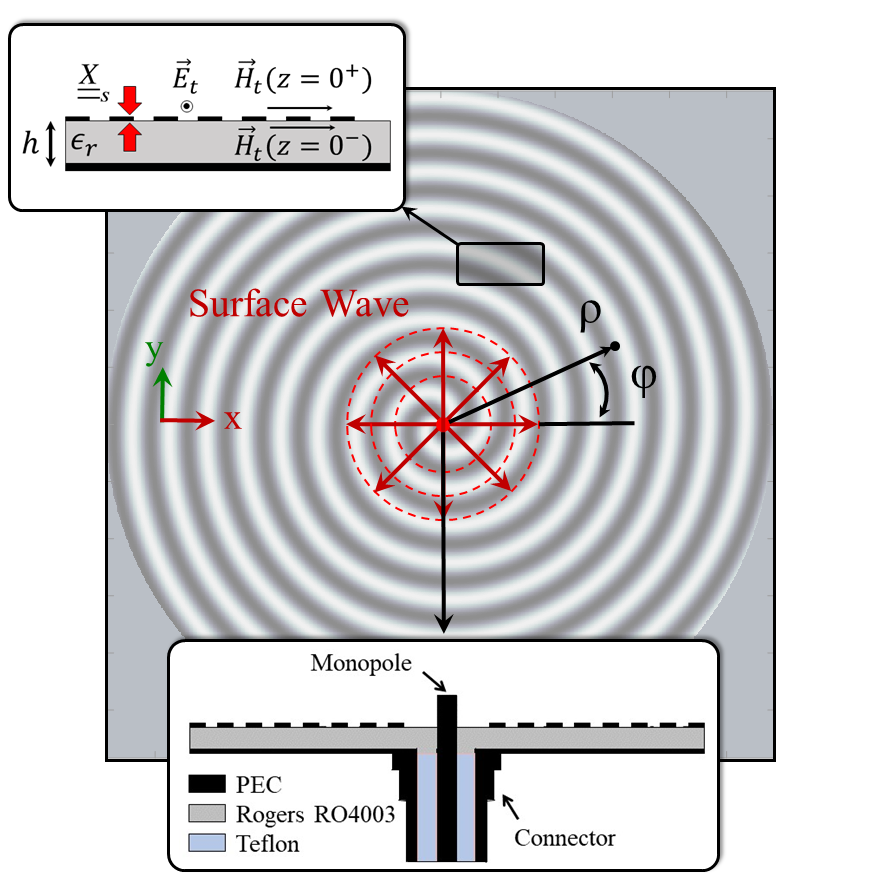}
      \caption{Conceptual schematic of a holographic antenna generating orbital angular momentum state in the leakage mode.
      The hologram is excited by a monopole launcher located at the center. The modulated surface impedance transforms the coupled surface wave into a leaky wave with arbitrary spin and orbital angular momentums.}
      \label{fig:fig1}
\end{figure}

\begin{figure}
      \centering
      \includegraphics[width = 0.4\textwidth]{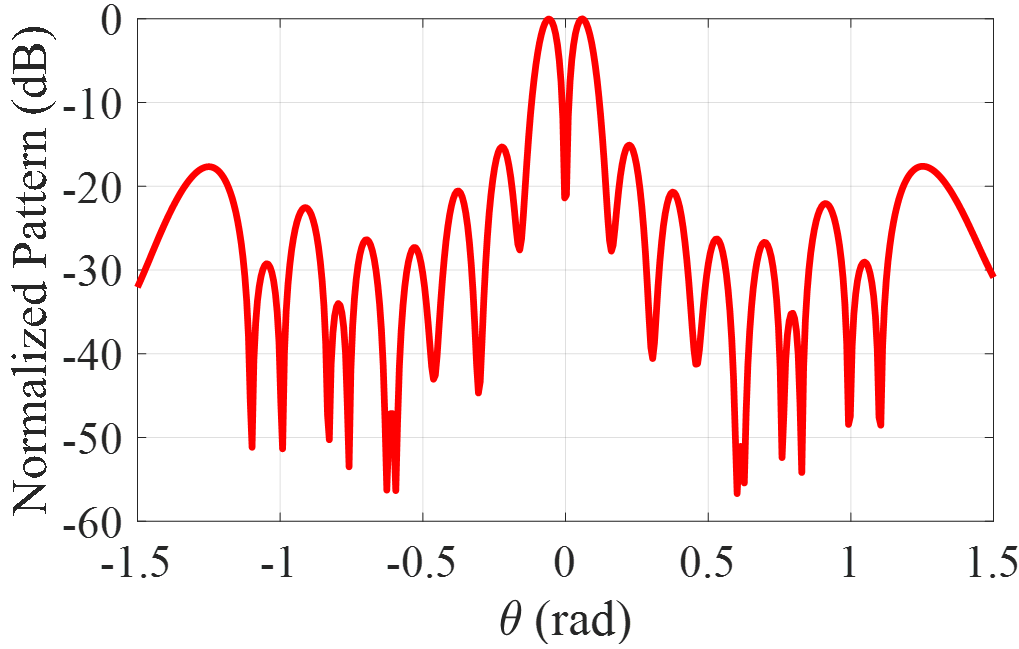}
      \caption{Two-dimensional radiation pattern for topological charge of $l = -2$.}
      \label{fig:dispersion_diagram}
\end{figure}

\begin{figure*}
	\centering
	\includegraphics[width = 1.0\textwidth]{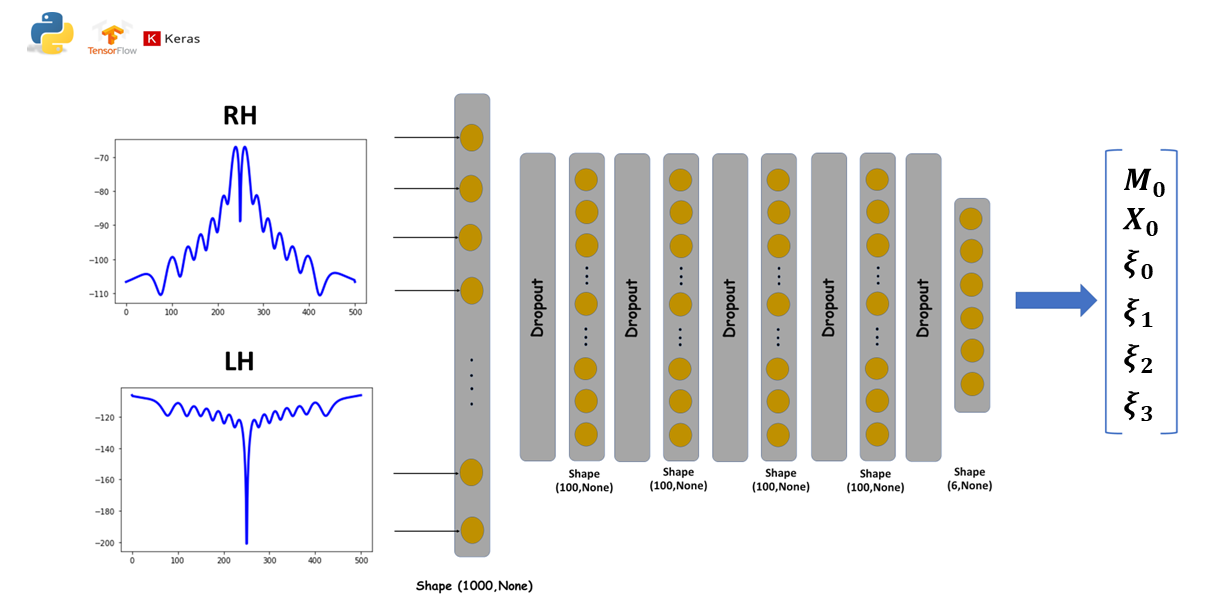}
	\caption{An overview of presented deep learning model.}
	\label{fig:DNN}
\end{figure*}

\section{Flat Optics and adiabatic Floquet-wave expansion method}
\subsection{Estimating aperture field}
An effective method for the generation of a polarized wave with the desired OAM spectrum is the holographic technique. 
Based on holography theory the following relation can be used to design and synthesize the surface impedance \cite{fong2010}:
\begin{equation}
	    Z_s(\vec{\rho}) = jX_0[ 1 + M(\hat{\rho})\times \Re \{\psi_{ref}(\vec{\rho}) \psi_{obj}^*(\vec{\rho})\}]
	    \label{eq:Z_s}
\end{equation}
where $\psi_{ref}(\hat{\rho})$ and $\psi_{obj}(\hat{\rho})$ are the reference wave (excited surface wave) and the objective wave (desired leaky-mode wave), respectively. 
$X_0$ is the average reactance controlling the slope of the surface wave dispersion and the antenna scanning rate. The coefficient $M(\hat{\rho})$ denotes the modulation index, which determines the leakage factor and radiation efficiency of the OAM generator.
The impedance equation defined in (\ref{eq:Z_s}) is valid for isotropic impedance boundary conditions. To generalize the holography theory for tensorial surface impedances, we may exploit the following equation \cite{minatti2015}:
\begin{equation}
	\underline{\underline{Z}}_s(\vec{\rho}) \cdot \hat{\rho} = jX_0 \eta_0[\vec{\rho} + 2 \Im \{\frac{\vec{E}_{ap}(\vec{\rho})}{-\hat{\phi} \cdot \vec{H}_t|_{z = 0^+}(\vec{\rho})}\}]
\end{equation}
In the above equation, $\vec{H}_t(\vec{\rho})$ is the tangential magnetic field at the upper limit of the interface being excited by the surface wave launcher
(as shown in Fig. \ref{fig:fig1}).
For center-fed holograms the excitation waveform may be  approximated as:
\begin{equation}
	\vec{H}_t(\vec{\rho}) -J_{sw} \approx H_1^{(2)}(\tilde{k}_0 \rho)
\end{equation}
The vector $\vec{E}_{ap}(\vec{\rho}) = E_{ax}(\vec{\rho}) \hat{x} + E_{ay}(\vec{\rho}) \hat{y}$ is the objective aperture field which must be estimated by the designer. According to the aperture field and stationary phase theory, the far-zone field may readily be obtained as follows \cite{balanis2005}:
\begin{equation}
	\vec{E}_{fz}(r, \theta, \phi) \approx \frac{jke^{-jkr}}{2\pi r} [F_\theta(\theta, \phi) \hat{\theta} + F_\phi(\theta, \phi)\hat{\phi}] 
	\label{eq:E_fz}
\end{equation}
\begin{equation}
	F_\theta(\theta, \phi) = \tilde{E}_{ax} \cos \phi + \tilde{E}_{ay} \sin \phi
\end{equation}
\begin{equation}
	F_\phi(\theta, \phi) = \cos \theta (-\tilde{E}_{ax} \sin \phi + \tilde{E}_{ay} \cos \phi)
\end{equation}
\begin{equation}
	\tilde{E}_{ax} = \iint_{ap} E_{ax}(\vec{\rho'}) e^{jk\rho' \sin \theta \cos(\phi - \phi')}\rho' d\rho' d\phi'
	\label{eq:E_ax_tilde}
\end{equation}
\begin{equation}
	\tilde{E}_{ay} = \iint_{ap} E_{ay}(\vec{\rho'}) e^{jk\rho' \sin \theta \cos(\phi - \phi')}\rho' d\rho' d\phi'
	\label{eq:E_ay_tilde}
\end{equation}
In (\ref{eq:Z_s}), the aperture field can be selected as desired. However, to ease the implementation and avoid the fabrication complexity, the following form may be appropriate for the aperture field components \cite{amini2021}:
\begin{equation}
	E_{ax, ay} (\vec{\rho}) = M_{x, y}(\vec{\rho}) \frac{E_0}{\sqrt{2\pi \rho \tilde{k}_0}} e^{-jk\rho \sin \theta_0 \cos (\phi - \phi_0)}
\end{equation}
The coefficients $M_x(\hat{\rho})$ and $M_y(\hat{\rho})$ as a function of $(\rho, \phi)$ indicate the modulation indices and their distribution are selected according to the vorticity and polarization states of the radiated wave. For instance, to have superimposed vorticity states in the OAM spectrum, the modulation indices can be defined as follows \cite{amini2022}:
\begin{equation}
      M_{x, y} (\vec{\rho})  = M_{0x, 0y}(\vec{\rho})\sum_{l = L_1}^{L_2} c_l e^{-jl \phi}
\end{equation}
where $l$ is the topological charge and $c_l$ determines the contribution of $l$'the harmonic in the OAM spectrum space.
Furthermore, the spin angular momentum (polarization)
state of the radiated wave prescribes the relationship between $M_x(\vec{\rho})$ and $M_y(\vec{\rho})$. In this paper, the radiated wave is assumed to be right-hand circularly polarized along the broadside direction. Therefore, we can write \cite{amini2021}:
\begin{equation}
      M_y(\vec{\rho}) = -j M_x(\vec{\rho})
\end{equation}
It is worth noting that, by selecting appropriate distributions for $M_{0x}(\vec{\rho})$ and $M_{0y}(\vec{\rho})$, the gain, null-depth, and side lobe level of the radiation pattern can be tuned. In this paper we use the following distributions to attain the goal:
\begin{equation}
	\begin{split}
		M_{0x, 0y} = M_0 \times[\xi_0 + \xi_1 \cos (\frac{\pi}{20  \lambda} \rho) + \\
		\xi_2 \cos (\frac{2\pi}{20  \lambda} \rho) + \xi_3 \cos (\frac{3 \pi}{20 \lambda} \rho)]
	\end{split}
\end{equation} 
In this work, we use the coefficient vector $\xi = [M_0, X_0, \xi_0, \xi_1, \xi_2, \xi_3]$ as the output sequence of the artificial neuron in order to achieve the desired far-zone pattern. 
\subsection{Floquet-wave expansion}
For quasi-periodic holographic structures, the precise form of current can be written in the form of:
\begin{equation}
	\vec{J}^{(n)} = \sum_n (J_\rho^{(n)} \hat{\rho} + J_\phi^{(n)} \hat{\phi}) e^{-jnKs(\vec{\rho})} H_1^{(2)} (\tilde{k}_0 \rho)
\end{equation}
for large arguments of the Hankel function (asymptotic form), the aforementioned equation can be simplified as:
\begin{equation}
	\vec{J}^{(n)} \approx \sum_n \sqrt{\frac{j2}{\pi \tilde{k}_0 \rho}}(J^{(n)}_\rho \hat{\rho} + J^{(n)}_\phi \hat{\phi}) e^{-j(nKs(\vec{\rho}) + \tilde{k}_0 \rho)}
\end{equation}
Therefore, the $n$-indexed phase and attenuation constants can be expressed as follows:
\begin{equation}
	\vec{\beta}^{(n)} = \Re \{\nabla_{\vec{\rho}}(nKs(\vec{\rho}) + \tilde{k}_0 \rho)\}  = \beta_{sw} + \delta_\beta + n\nabla_{\vec{\rho}}(Ks(\vec{\rho}))
\end{equation} 
\begin{equation}
	\vec{\alpha}(\vec{\rho}) = \Im\{\nabla_{\vec{\rho}}(nKs(\vec{\rho}) + \tilde{k}_0 \rho)\} = \Im \{\tilde{k}_0 \rho\}
\end{equation}
where $\nabla_{\vec{\rho}}$ denoted gradient operator applied on tangential components. It is worth noting that, $\alpha(\vec{\rho})$ is independent of $n$ due to the fact that the modulation phase ($Ks(\vec{\rho})$) is always real.
The parameter $\tilde{k}_0$ is the wavenumber of the fundamental mode whose value must be determined accurately. A robust method to calculate the wavenumber is the Flat Optics method proposed in 
\cite{minatti2016}.
In this method, the eigenvalue equation for fundamental mode is established and by setting the determinant of the coefficient matrix equal to zero, eigenstates can be accurately calculated.
Consequently, the corresponding dispersion curves and eigenvectors are obtained. 
Fig. \ref{fig:dispersion_diagram}
shows the right-hand radiation pattern of the proposed antenna generating OAM mode with a topological charge of $l = -2$ at the broadside. 
The coefficient vector $\xi$ is assumed to be $\xi = [0.05, 0.5,0.05, 0, 0, 0]$ and the operating frequency is chosen as  18 GHz.  
Rogers RO4003 with $\epsilon_r = 3.55$ and thickness $h = 1.524$ mm is used as the dielectric host medium. The modulated impedance cladding covers the substrate to excite the desired angular momentums (both spin and orbital angular momentums).

Using deep learning, it is possible to design machines that adapt to their environment based on both their own knowledge and experience to shape the radiated pattern. To be precise, deep learning attempts to create a machine that is able to learn and operate without direct guidance as opposed to explicitly planning and dictating the actions. Instead of programming each operation, deep learning feeds its logic from data into an algorithm, and then the algorithm learns its logic from the data 
\cite{lecun2015,krizhevsky2017}.

\section{Dataset and deep learning}
Since there is no clear trend in changing the values of the coefficient components $\xi_i$, i.e. the elements of the coefficient vector $\xi$, a special procedure should be explored to control the surface impedance of OAM-carrying HLWA. To do this, it is required to generate a dataset of mentioned components which change as follows:
\begin{equation}
	\begin{split}
		0.1 \leq X_0 \leq 1.1 \\
		0.1 \leq M_0 \leq 1.0 \\
		0 \leq \xi_{0, 1, 2} \leq 0.8 \\
		0 \leq \xi_3 \leq 1.0 
	\end{split}
\end{equation} 
The corresponding intervals are chosen in such a way that the synthesized impedance could be easily implemented. In this vein, by utilizing Matlab software a total of 77,000 data-set have been generated in this study, of which $70\%$ are used for training, $15\%$ for validation, and $15\%$ for testing. In order to accomplish this goal, the following DNN architecture has been designed. The DNN algorithm is implemented using Python, Tensorflow version 2.2.0, and Keras version 2.3.1. Two layers are used sequentially in this model, dense and dropout. The DNN is designed using linear activation and Relu activation function. The batch size and learning rate for the model are set at 25 and 1e-6. In order to tune the weighting values, ADAM optimization is employed, and the Mean Squared Error is used as the loss function, as follows:
\begin{align}
	Loss = \frac{1}{N}\displaystyle\sum\limits_{i=1}^N (y_{i}-\hat{y_{i}})^2
    \label{eq:Loss}
\end{align}
Which y is the actual value, $\hat{y}$ is the predicted value, and the number of data points is defined as $N$.
Fig. \ref{fig:DNN} illustrates the DNN architecture in which both Right-Hand (RH) and Left-Hand (LH) polarized radiation patterns of OAM waves as input have been imported into the proposed network. In this network, contrary to other traditional methods such as Genetic Algorithm (GA), both RH and LH radiation patterns can be controlled, simultaneously. Also, there is no need to design a new network for each new pattern and the provided network is responsive to any desired radiation pattern. The output matrix denotes the coefficient components of $\xi$ which are estimated by the proposed DNN. 

\begin{figure}
	\centering
	\includegraphics[width = 0.45\textwidth]{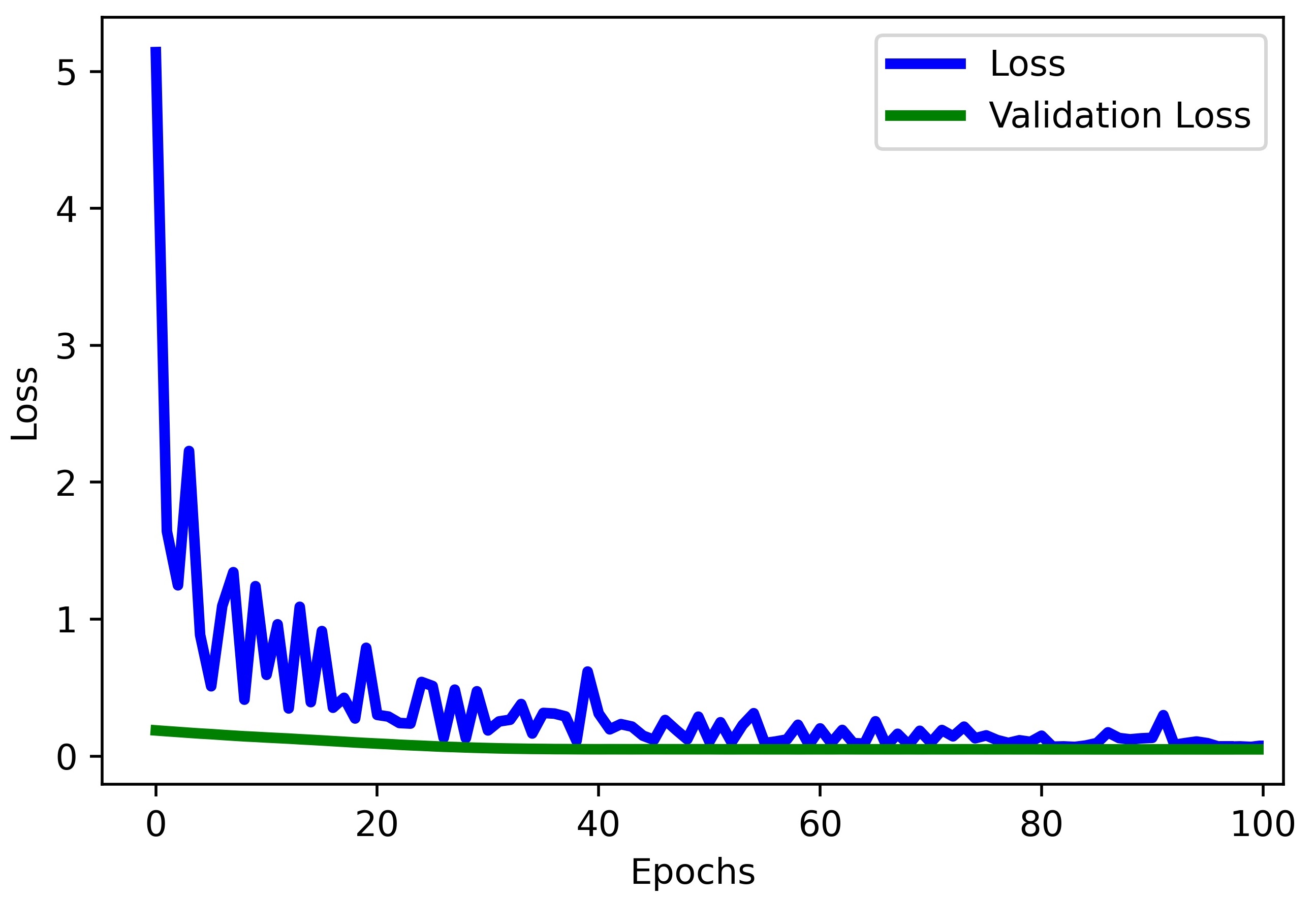}
	\caption{Curves of loss function relative to 100 Epochs.}
	\label{fig:loss}
\end{figure}

By employing (\ref{eq:Loss}) loss function curves have been drawn. As demonstrated in Fig. \ref{fig:loss}, the proposed DNN has a low loss which refers to high precision. 

\begin{figure}
	\centering
	\subfloat[]{\includegraphics[width=0.4\textwidth]{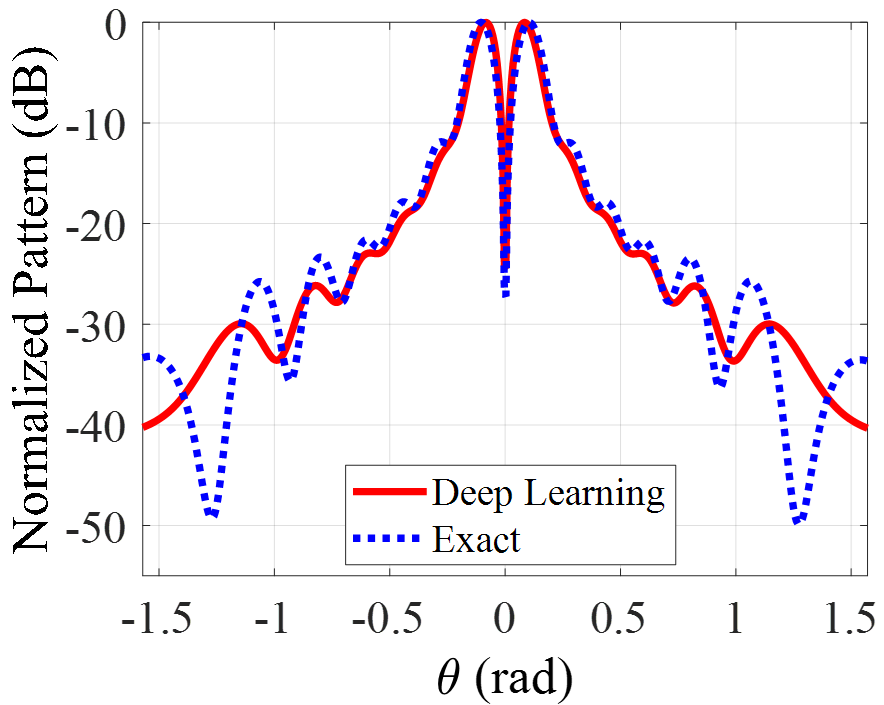}}
	\qquad
	\subfloat[]{\includegraphics[width=0.4\textwidth]{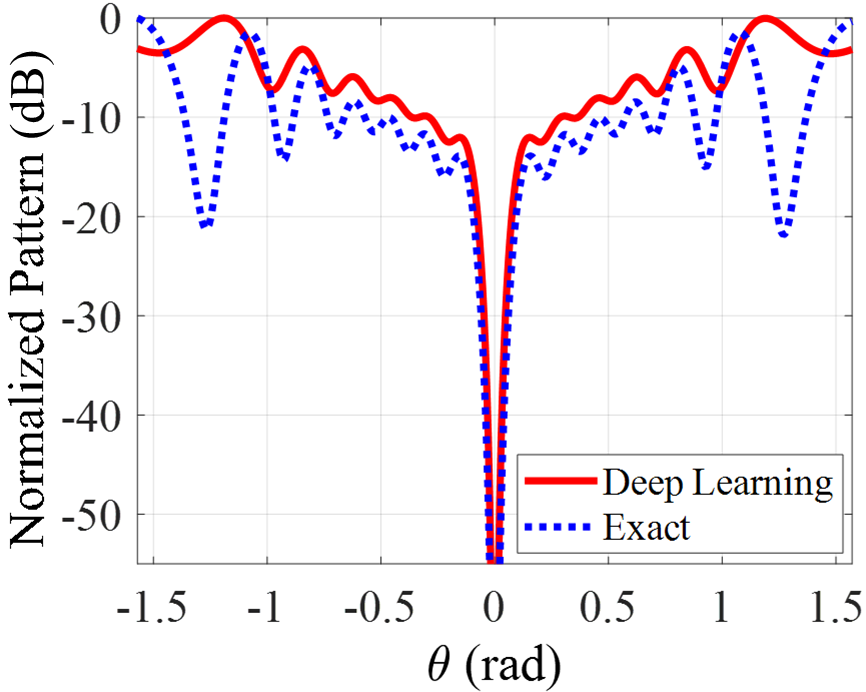}}
	\qquad
	\caption{The (a) right-hand and (b) left-hand patterns generated by deep learning are compared with the actual patterns. The coefficient vector $[M_0, X_0, \xi_0, \xi_1, \xi_2, \xi_3]$ has values of [0.4, 0.5, 0.0, 0.5, 0.2, 0.3] and [0.537, 0.541, 0.195, 0.203, 0.195, 0.204] for exact and estimated states, respectively.}
	\label{fig:lchp_rhcp}
\end{figure}

\begin{figure}
	\centering
	\subfloat[]{\includegraphics[width=0.37\textwidth]{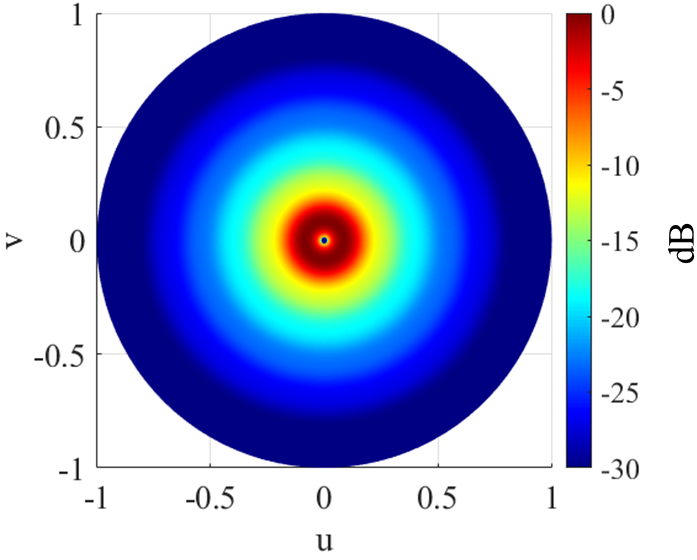}}
	\qquad
	\subfloat[]{\includegraphics[width=0.37\textwidth]{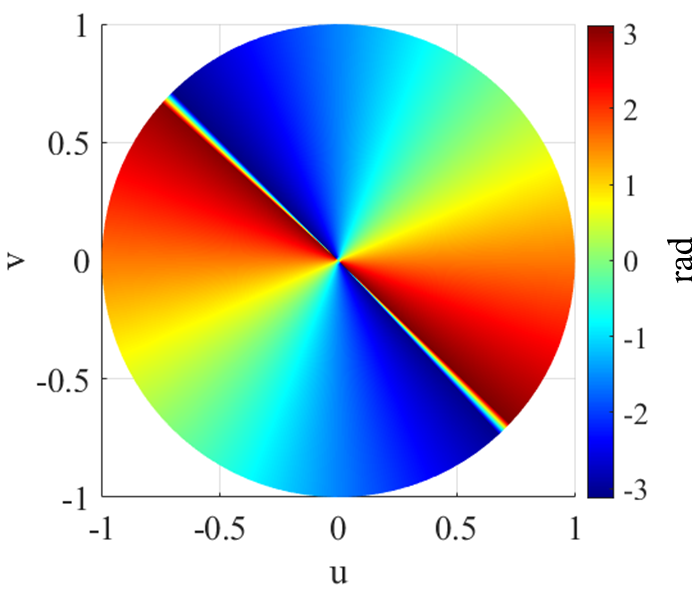}}
	\qquad
	\caption{Analyrical results of right-hand component for the proposed hologram, (a) magnitude and (b) phase distribution in the u-v plane.}
	\label{fig:lchp_rhcp_2d}
\end{figure}

In order to weigh the speed of the network, we presented the training network time, the inference time, the size of deep learning models, and the number of parameters for deep neural networks all in Table \ref{tab:tab1}. This study uses Google Colab and a Tesla T4 GPU with 15 GB of RAM. However, with respect to Table \ref{tab:tab1} it can be realized that the trained network is of high speed. Moreover, the comparison results between exact and deep learning solutions are given in Fig. \ref{fig:lchp_rhcp}a and b. As it is clear, the estimated results of the addressed network are in good agreement with the exact functions for both RH and LH radiation patterns. Graphically, the magnitude and phase of far-zone patterns corresponding to the Fig. \ref{fig:lchp_rhcp}a are plotted in Fig. \ref{fig:lchp_rhcp_2d}a and b, respectively. It can be inferred that the exemplified radiation pattern has a low SLL and deep central null accompanied by a right-handed circular polarization ability indicating the vorticity state of the radiated beam.
\renewcommand{\arraystretch}{1.5}
\begin{table}[]
	\centering
	\caption{An indicator of the number of parameters in the network, the size of the model, training time, and the inference time for one input.}
	\centering
	\begin{tabular}{|c|c|}
		\hline
		Training Time       & 15.3 Min  \\ \hline
		Inference Time      & 0.082 Sec \\ \hline
		Model Size          & 13 MB     \\ \hline
		Trainable Parameters & 1,132,006 \\ \hline
	\end{tabular}
	\label{tab:tab1}
\end{table}
\section{Conclusion}
In this paper, an analytical method based on the FO and deep learning framework is proposed to design and analyze leaky-wave vortex beam generators. In general, it can be extremely challenging and time-consuming to mathematically determine the correct values for geometrical parameters in metasurface-based leaky-wave antennas. In light of the complex and labor-intensive process of determining the correct parameter values using mathematical techniques, machine learning techniques can provide a more effective and efficient solution.

\ifCLASSOPTIONcaptionsoff
  \newpage
\fi



%\newpage


%\vfill


\end{document}